\documentclass[5p, authoryear]{elsarticle}

\usepackage{hyperref}
\usepackage{amssymb,amsmath,amsfonts,graphicx,aas_macros, times, natbib}
\usepackage{algorithm, algpseudocode}

\newcommand{\hfof}{{\sc hfof}}

\newcommand{\kdcount}{{\href{http://rainwoodman.github.io/kdcount}{kdcount}}}
\newcommand{\kdtree}{{$k$-d tree}}
\newcommand{\kdtrees}{{$k$-d trees\ }}
\DeclareMathOperator{\xor}{xor}

\defcitealias{Sinha_2017}{Corrfunc (Sinha and Garrison, in prep.)} 

\journal{Astronomy and Computing}
\begin{document}
\begin{frontmatter}
\title{Tree-less 3d Friends-of-Friends using Spatial Hashing}

\author[ucr,corauth]{Peter Creasey}
\ead{peter.creasey@ucr.edu}

\address[ucr]{Department of Physics and Astronomy, University of California, Riverside, California 92507, USA}

\begin{abstract} 
  I describe a fast algorithm for the identification of connected 
  sets of points where the point-wise connections are determined by a fixed spatial 
  distance - a task commonly referred to in the cosmological simulation community
  as Friends-of-Friends (FOF) group finding. This technique sorts particles into fine 
  cells sufficiently compact to guarantee their cohabitants are linked, and uses 
  locality sensitive hashing to search for neighbouring (blocks of) cells.
  Tests on N-body simulations of up to a billion particles exhibit speed 
  increases of factors up to $20\times$ compared with FOF via trees (a factor 
  around $8$ is typical),
  and are consistently complete in less than the time of a \kdtree\  
  construction, giving it an intrinsic advantage over tree-based 
  methods. The code is open-source and available online at 
  \url{https://github.com/pec27/hfof}.
\end{abstract}


\end{frontmatter}

\section{Introduction}
\label{sec:intro}
A way to identify dense groups of points in $\mathbb{R}^k$ is to construct connected components of points where direct connections are given for all pairs of points whose Euclidean separation is less than a `linking length' $b$. 
This task is particularly common when processing cosmological simulations of the $\Lambda$ cold-dark matter model to find the statistics of halos, 
which are virialised objects with a mean density of approximately $200\times$ the critical density of the universe \citep{Gunn_1972,Bertschinger_1985,Eke_1996}.
Simulations of these objects discretise the (primarily dark) matter distribution into $N$ bodies \citep{Davis_1985}, and at any given time-scale of interest
a catalogue of the connected components (or `friends-of-friends' groups) in $\mathbb{R}^3$ is constructed (e.g. \citealp{Jenkins_2001, Reed_2003, Reed_2007, Crocce_2010, Courtin_2011, Angulo_2012},
although other alternatives exist, see \citealp{Knebe_2011} for an overview).
The data sets of these simulations have grown from 32,768 particles \citep{Davis_1985} to the trillions of particles this decade \citep{Skillman_2014}, making the production of these group catalogues challenging.

The ubiquitous algorithm for finding these friends-of-friends (hereafter FOF) groups is to perform a breadth-first search \citep[e.g.][]{Huchra_1982}.
In this algorithm, finding connected components proceeds in the following manner: a stack of boundary points is maintained (initialised with a single point), and at each step a point is removed (marked as linked) and replaced by all its (unlinked) neighbours within the linking length, and this proceeds until the stack is empty, and the component is complete.
This fixed-radius neighbour search is performed via organisation of the points into a \kdtree, a binary space partitioning structure where neighbour searches can be performed in $O(\log n)$ operations, $n$ being the total number of points.
Examples of such codes include \citet{Behroozi_2013a}, the FOF code from the NbodyShop\footnote{\url{http://faculty.washington.edu/trq/hpcc/}, see also \url{https://github.com/N-BodyShop/fof} for the code}, which is the almost unmodified ancestor of more recent codes such as \href{https://github.com/junkoda/cola_halo}{Cola} \citep{Koda_2016,Carter_2018}, {\sc yt}\ \citep{Turk_2011} and probably many others unknown to this author. 
As far as I am aware \kdtrees are used to perform the neighbour finding step in the non-public codes also, such as \citet{Kwon_2010, Fu_2010} and {\sc Arepo}\ \citep[][and also the non-public version of its predecessor {\sc Gadget}-2]{Arepo}.
Some of these codes have been designed to create the group catalogue in parallel (often on the same cluster as the simulation), to mitigate the analysis problems.

Recently \citet{Feng_2017} have released an open source ($k$-dimensional) FOF algorithm \kdcount\footnote{see \url{http://rainwoodman.github.io/kdcount}} that is used in {\sc nbodykit}\  \citep{Nbodykit}. This algorithm uses the dual tree method \citep[e.g.][]{Moore_2001} which exploits the fact that the searching points are hierarchically organised, allowing neighbour calculations (either inclusions or exclusions) to be calculated (typically excluded) for entire branches of the search tree.
Their algorithm is not strictly breadth-first, a consequence of which is the need to merge components using a (customised) disjoint-set algorithm \citep{Tarjan_1975}. 

An alternative method for neighbour searches is the mapping of points on a fixed-grid, for example in the `chaining mesh' method of \citet[sec 8.4.1]{Hockney_1988} for a short-range component of the Coulomb force,
and in the correlation function code Corrfunc (ascl:1703.003).
By choosing a cell width greater than the search radius, one guarantees that all neighbours are within the 26 adjacent cells (in 3-d). 
Since the extent of the short range force is generally a multiple of the interparticle separation, this mesh is coarse w.r.t. the particles, corresponding to a modest memory footprint. Unfortunately, in the application FOF one is generally interested in linking lengths of $0.2\times$ the interparticle spacing \citep[e.g.][]{Davis_1985}, implying meshes of (at least) $125\times$ the particle count, and correspondingly a prohibitively large memory footprint. 

A method to avoid such large data structures is to store only the filled cells, mapping them into a 1-d hash-table \citep{Yuval_1975, Bentley_1979} such that neighbouring cells can be (speculatively) searched at the map (hash) of their location.
Such a method has been employed for fixed-radius neighbour searches \citep[e.g.][sometimes referred to as locality sensitive hashing]{Teschner_2003, Hastings_2005}. 
This is $O(1)$ for look-ups, though limitations include the expense of the hash function, the cost of resolving collisions (cells mapped to the same index) and the decreased coherence of memory accesses.

Spatial hashing has been successfully implemented by \citet{Wu_2007} and \citet{Vijayalaksmi_2012} for the related clustering algorithm DBSCAN, which is a generalisation of FOF to connecting components only about a subset of `core' points \citep{Ester_96}. 

In practice these codes are not applied to FOF calculations, possibly because they have not been optimised for this specialised use-case.
Spatial hashing appears to be less common in computational physics, with exceptions such as the parallelisation scheme of \citet{Warren_1993} and in the level set tracking methods of \citet{Brun_2012}.

This paper describes a novel algorithm for performing FOF in 3-d by grouping points into fine mesh whose cells are sufficiently compact to guarantee their points will be connected. These filled cells are grouped into $4^3$ blocks which are stored via spatial hashing, the use of blocks decreasing the average number of hash-look-ups per filled cell. 
The merging of cells happens `on the fly' as the blocks are inserted in a raster order, i.e. neighbours queries are only performed over blocks previously inserted in the table, and then the components are connected via the disjoint-sets algorithm, in a manner similar to \citet{Feng_2017}. 
An example implementation is provided at \url{https://github.com/pec27/hfof}.

This paper is organised as follows. Section~\ref{sec:method} describes the spatial hashing and linking algorithm, optimisations, and the method applied for periodic domains. Section~\ref{sec:comp} describes the comparison codes and test sets. Section~\ref{sec:results} analyses the performance and compares with other codes and Section~\ref{sec:conc} concludes.

\section{Spatial hashing for fixed-distance neighbour linking}\label{sec:method}

In this section a methodology for FOF group finding via spatial hashing is described. 
Whilst this algorithm is not limited to cosmological simulations, these are the motivation, and some consideration of their features for this purpose as described in Sec.~\ref{sec:cosmo}.
Sec.~\ref{sec:cells} describes the arrangement of points into cells compact enough to guarantee connectivity, and their aggregation into blocks to reduce the number of lookups. Sec.~\ref{sec:hash_func} describes the hash function and \ref{sec:periodicity} describes the adjustments to account for periodic domains.

\subsection{Matter distribution in cosmological simulations}\label{sec:cosmo}

In the cosmological context, the clustering of matter produces halos which at the low-mass regime have a mass function approximating a power law
\begin{equation}\label{eq:mass_func}
\frac{{\rm d}N}{{\rm d} M} \propto M^{-\alpha}
\end{equation}
with $\alpha \approx 1.9$ (e.g. \citealp{Reed_2007}), and notably $\alpha>1$ implies a divergent low-mass tail,
 i.e. there should be an infinite number density of low-mass clusters, our discrimination of them limited only by our finite mass resolution (this is not strictly speaking true of the real universe, where diffusion damping terms will limit very low mass halos, but these are rarely resolved in cosmological simulations). A corollary of this is that the groups found are likely to be dominated (by number) by single particle groups\footnote{in the analysis of cosmological simulations groups with small (e.g. $<20$) particles are generally ignored, but at the stage of constructing FOF groups these have yet to be filtered}, and also that the number of groups is a significant fraction of the total number of particles (typically around one-third for cosmological simulations). 
As such a FOF algorithm needs to be efficient in the cases where the neighbourhood within a linking length is empty.

At the other extreme is that of high mass groups. Given the previous paragraph it may be tempting to think that most points are in small groups, however this is not the case. 
This can also be seen from Eqn.~(\ref{eq:mass_func}), since $\int M {\rm d}M / \int {\rm d}M$ (i.e. the mass-weighted average halo mass) would have a divergent high-mass contribution, i.e. the average particle is in a group of $\gg 1$ particles, the exact number depending upon the mass function to higher masses (which in discrete simulations often depends upon artificial limitations such as the box size). 
As such the linking component of a FOF algorithm needs to scale well, in order to handle the connection of points to large groups.

Whilst both of these extremes need to be handled by group finding algorithms, I find in general the former seems more demanding, in that a significant fraction of the particles have zero neighbours within the linking length, and the majority of the computational time is spent confirming that these particles are truly isolated (see for example the 2nd panel of Fig.~\ref{fig:adj_idx}). It is helpful to keep this in mind during the following section.

\subsection{Cell and block organisation}\label{sec:cells}
At the finest level, each particle is assigned to a cell according to its position in a lattice with cell-width
\begin{equation}
c = \frac{b}{\sqrt{3}}
\end{equation}
where $b$ is the linking length. Since the maximum distance between vertices in a unit hypercube in $\mathbb{R}^k$ is $\sqrt{k}$, this guarantees that any points in the same cell must belong to the same FOF group, which essentially reduces the problem of linking points to one of linking cells, and hereafter I will almost exclusively talk in terms of cells. The filled cells are sorted in raster order (i.e. sorted by $z$ then $y$ then $x$), which immediately places a bound on the the complexity of the algorithm to be at least $O(n\log n)$, similar to that of the \kdtree\ construction.

\begin{figure*}
\includegraphics[width=2\columnwidth]{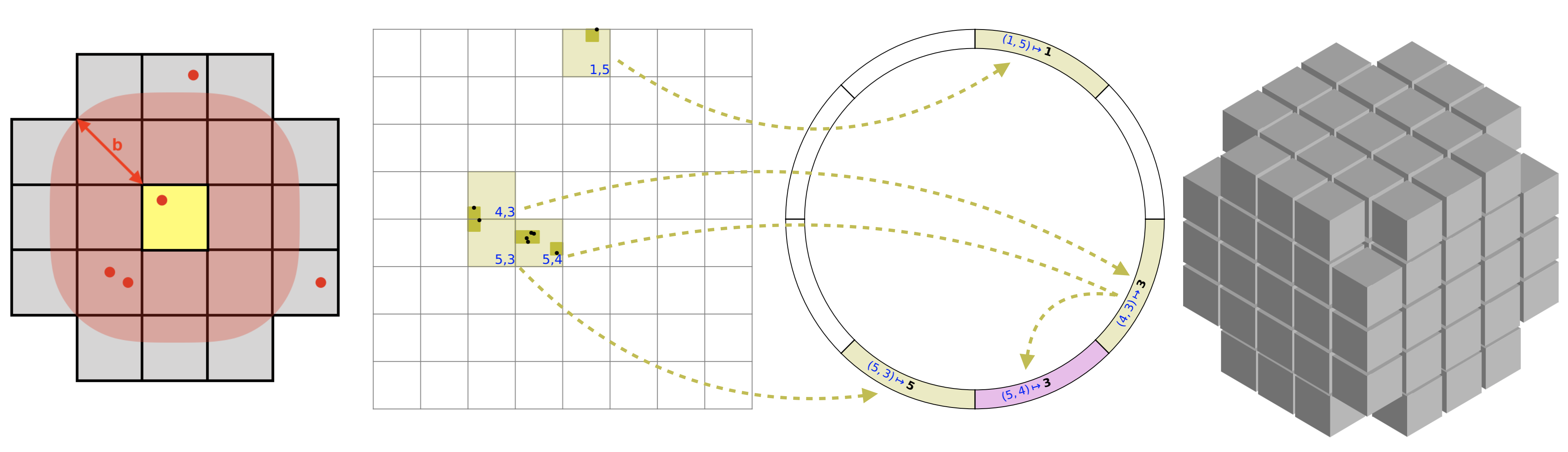}
\caption{2 and 3-dimensional representation of the method, from left to right. \emph{Far-left}, a filled cell (\emph{yellow square}) and the neighbouring 2-dimensional `stencil' (\emph{grey squares}) of 20 neighbouring cells which could contain points within $\sqrt{2}$ cell widths (the linking length $b$ denoted with the \emph{red arrow}), where the exact locus for the \emph{yellow square} is given by the \emph{red shaded region} and some example points are given by \emph{filled red dots}. \emph{Centre-left}, points from a (tiny slice of a) real simulation (\emph{black dots}) which are grouped into filled cells (\emph{dark yellow squares}) which are themselves grouped into $4^2$ blocks (\emph{light yellow squares}). These blocks are then mapped (\emph{light yellow arrows}) to the hash-table in the \emph{centre-right panel} at entries (\emph{light yellow segments}) whose positions that are hashes modulo $2^3$ (values in \emph{black}) of their spatial indices (in \emph{blue}), with collisions demoted to the next available position (\emph{magenta}). \emph{Far-right} the 3-d stencil of the (116) neighbouring cells that can be with $\sqrt{3}$ cell widths of the innermost cell.}
\label{fig:adj_idx}
\end{figure*}

This relationship of cell size to linking length is illustrated in Fig.~\ref{fig:adj_idx} (first panel), where the locus of potential neighbours for positions in the central cell is highlighted. 
This lattice size guarantees that any neighbouring particle within a distance $<b$ must be within a `stencil' of the 116 adjacent cells (Fig.~\ref{fig:adj_idx} rightmost panel). This can be reduced by a factor of 2, to 58 neighbouring cells, by assuming that the lattice is built in raster order and using the symmetry of the distance metric\footnote{i.e. 58 subsequent cells will be connected when they search for the current cell}, however 58 turns out to be a rather large number of neighbour searches per cell (see discussion in Sec.~\ref{sec:hash_func} about optimisation of hash-table look-ups). As such, the cells are grouped into \emph{blocks} of $4\times4\times4$ (i.e. 64) cells, i.e. the block width is
\begin{equation}
\Delta = 4c
\end{equation}
(blocks of $4^3$ are assumed for the remainder of this paper, with the exception of the testing done in Section~\ref{sec:results}) and the block indices $i,j,k$ of each point $x,y,z$ are given by
\begin{equation}
i = \lfloor \frac{x - \min x}\Delta \rfloor , j = \lfloor \frac{y - \min y}\Delta \rfloor , k = \lfloor \frac{z - \min z}\Delta \rfloor \, .
\end{equation}

Neighbouring filled cells are thus guaranteed to be within the $3^3-1=26$ adjacent blocks, and again building the blocks in raster order allows the exploitation of symmetry to reduce the searches to $S=13$. 
The choice to search only for those blocks with a smaller raster index also means we can link them `on the fly' with their insertion into the hash-table (see Sec.~\ref{sec:hash_func}). 
An additional step is made by pre-computing the overlap of the adjacent blocks that must be considered per cell (i.e. cells near the corners of the block under consideration are only within a distance $b$ of a subset of the $S$ neighbouring blocks). Counting which of the $S$ blocks may contain a direct neighbour (i.e. applying the stencil in the rightmost panel of Fig.~\ref{fig:adj_idx} for each of the $64$ cells) reveals $220$ blocks, or an average of $\frac{55}{16}\approx 3.4$.

The average number of filled cells per block is often quite low, and seems to be around $2$ for cosmological simulations, with a weak dependence on resolution (see the values in Table~\ref{tab:simtable} in section~\ref{sec:test_sets}, which range from around $1.7$ for the largest particle mass to $2.4$ for the smallest). I suspect this near scale-invariance is due to the power-law low-mass tail of the halo mass function, with a distribution similar to that of a negative binomial \citep[see e.g.][]{Ata_2015}. 
On the other hand, if the data is nearly 2-d such as the galaxy in sec.~\ref{sec:test_sets} then it is much higher at $\approx 15$.

To account for the low volume-filling values the filled cells are simply arranged in a ragged array, i.e. an array of filled cells per block, whose length varies per number of (filled) cells in the block (hereafter for brevity all cells are assumed to be filled unless stated otherwise). This is in some sense a nested `chaining mesh' \citep[page 277]{Hockney_1988}, in that each block points to its first filled cell, and the cells to their first particles.
Each cell knows its physical position (of $4^3$) in the block, these being used to test against a stencil (as per the right panel of Fig.~\ref{fig:adj_idx}) of which (cells in) neighbouring blocks are within a distance $b$. 
The rather laborious task of evaluating masks of which cell (of 64) in each block (of up to $1+S=14$ neighbouring inserted blocks and the current block) could contain a point within a distance $b$ of the current cell (of 64) is precomputed.

In order to compare cells for linking, we require algorithms to
\begin{enumerate}
\item Determine whether the two cells are already in the same set, i.e. no pairwise point comparisons are required since they are already linked, and,
\item If pairwise point comparisons have been performed and a connected pair found, the components need to be merged.
\end{enumerate}
It is relatively easy to `home-brew' an algorithm that is $O(1)$ in one of these operations but linear in the size of the set for the other\footnote{e.g. a linked list is $O(1)$ to connect, but linear to determine the root. Conversely a linked list with a root pointer is $O(1)$ to determine the root, but linear to update all those root pointers. This author of course had to implement both before educating himself.}, the latter coming to dominate for large point sets. Fortunately a rapid (extremely sub-logarithmic) algorithm to do both exists, known as the disjoint-sets algorithm. This algorithm has been comprehensively discussed elsewhere (see \citealp{Tarjan_1975} or a more modern discussion in \citealp{Cormen_2007}), and as such the following only describes the essential details for this context.

The disjoint-sets structure forms a tree for each set, requiring that each cell keeps track of two values, being
\begin{itemize}
\item \emph{Parent:} An index of the parent (or to itself, if the root), which itself may have a parent, and a
\item \emph{Rank:} approximately determining the depth of the tree, initialised to zero for isolated cells.
\end{itemize}
The operations 1 and 2 are then performed as
\begin{enumerate}
\item Root comparison using path compression: The root for each cell is determined by following the parent indices until a root is found. After each walk the intermediate nodes have their parents set to the root (known as `path compression', a process which keeps the trees shallow).
\item Linking (union by rank): The set with the lower rank root is inserted to that of the higher, by setting the parent index of the lower rank root to the higher. If the two roots have the same rank then choose arbitrarily (in the implementation the adjacent is inserted into the current) and increment the rank of the root.
\end{enumerate}
It turns out (see \citealp{Cormen_2007}) that these trees remain sufficiently shallow (i.e. few steps to the root) that both operations are sub-logarithmic with a small constant factor. 

\subsection{Hash function}\label{sec:hash_func}

For a uniform cosmological simulation of $N$ particles in a box of size $L$, a fiducial linking length of $b=\frac{0.2L}{N^{1/3}}$ corresponds to lattices with a total number of (empty and filled) blocks
\begin{equation}
\frac{L^3}{\Delta^3} = \left( \frac{\sqrt{3}}{4\times0.2} \right)^3 N \approx 10.1 N \, ,\nonumber
\end{equation}
which for large simulations (where the particle data can barely be fit into memory) is clearly problematic. This problem can be essentially eliminated by the process of spatial hashing, which avoids expensive binary searches (for neighbours) or large memory usage by storing only the \emph{filled} blocks in a table where the insertion indices correspond to the hashes of the spatial indices (as shown in Fig.~\ref{fig:adj_idx}, middle panels). Collisions (overlapping indices) are resolved by promotion to the subsequent entry (modulo the hash table size, usually referred to as `open addressing' with linear probing), and the optimal fraction $\lambda$ of filled entries (often referred to as the `load') is usually around 60-70\%, taking into account cache misses (due to large table size) and collisions, the latter ideally occurring at a random rate but in practice some occur to poor hashing (i.e the function does not truly de-cluster the data).

Spatial hashing is a specialised version of hashing in that the number of hashes performed relative to the amount of data is relatively large, since the neighbouring volume grows exponentially with the number of dimensions, 
and that the searches are `speculative', meaning the search is performed without knowledge of whether there is a filled block at the requested location\footnote{Queries in conventional hash-tables are usually for data that is known to exist.}.
As such we are interested in hash functions which are relatively fast over more complex hash functions that have superior de-clustering.

I have chosen a scheme where each block is assigned an index $\Phi$, where
\begin{equation}\label{eq:block_index}
\Phi(i,j,k) = P_{\rm x} i + P_{\rm y} j + k
\end{equation}
with prime numbers $P_{\rm x}$ and $P_{\rm y}$ chosen such that $P_{\rm y}> \frac{L}\Delta + 1$ and $P_{\rm x}> \frac{L  P_{\rm y}}\Delta +1$ (i.e. ordering by $\Phi$ is still ordering by $i$ then $j$ then $k$, the additional $+1$s being required to `buffer' the rightmost values from the next row), and the 
primes are found from the Miller-Rabin test \citep{Rabin_1980} as the smallest values satisfying these inequalities. 
A 64-bit integer is used to store $\Phi$, since this safely covers $\frac{L}\Delta$ up to approximately 2,600,000, or $N<10^{18}$ particles (for $b=0.2$), which is still well out of reach of all current cosmological simulations. 
The choice of hash function is given by
\begin{equation}\label{eq:hash}
H(i,j,k) = \Phi(i,j,k) \times Q \mod 2^n
\end{equation}
where $2^n$ is the size of the hash table and $Q$ is a prime number of similar magnitude to $2^n$. 

This choice of Eqns.~(\ref{eq:block_index}) \& (\ref{eq:hash}) has the convenient property that the relative hashes of adjacent blocks are 
straightforward to compute as
\begin{eqnarray}
H(i+1,j,k) - H(i,j,k) =& QP_{\rm x} & \mod 2^n , \nonumber \\
H(i,j+1,k) - H(i,j,k) =& QP_{\rm y} &\mod 2^n , \nonumber \\
H(i,j,k+1) - H(i,j,k) =& Q & \mod 2^n , \nonumber  \\
\dots && \nonumber
\end{eqnarray}
with all adjacent blocks following from substitution of indices into Eqns.~(\ref{eq:block_index})~\&~(\ref{eq:hash}).

Whilst the use of a multiplicative hash in Eqn.~(\ref{eq:hash}) is easily understandable in terms of simplicity, the use of prime multipliers in Eqn.~(\ref{eq:block_index}) for conversion from a 3-d index to 1-d may not be immediately clear. 
A simplification of Eqn.~(\ref{eq:block_index}) would be to omit the requirement that the $P_{\rm x}$ and $P_{\rm y}$ multipliers be prime, however this causes problems in the cases when one these values is a multiple of a power of 2. In particular if it is a multiple of $2^m$ (for some nonzero $m$) then it is no-longer co-prime with $2^n$ and the hash function outputs identical values for any increment that is a multiple of $2^{n-m}$ along the corresponding axis (or if one is thinking in terms of bit-wise representations the increments do not affect the lowest $m$ bits of the hash). This use of prime multipliers is similar to \citet{Teschner_2003} in which indices were combined as $P_1 i \xor P_2 j \xor P_3 k$, however the use of the $\xor$-function in that work complicates the evaluation of hashes for adjacent indices (i.e. the relative hashes are no-longer independent of $i$,$j$,$k$) because $\xor$ is not distributive over addition. 

Notably building the hash table incrementally (as one adds blocks ordered by $\Phi$) turns out to have the additional benefit that the speculative neighbour searches 
are performed when the hash table is only partially full. Taking the continuous approximation for random hashing that the expected rate of collisions (i.e. false elements found) when inserting/searching a single element into a table of fill $\lambda$ is $F=\frac{\lambda}{1-\lambda}$ (i.e. the geometric sum of $\lambda^n$), the \emph{average} collision rate (as the table is incrementally built) is given by the convex function
\begin{equation}
\bar{F}(\lambda) = \frac{1}{\lambda}\int_0^\lambda  \frac{s \, {\rm d}s}{1-s}= -1 - \frac{\log\, 1-\lambda}{\lambda}
\end{equation}
where $\log$ refers to the natural logarithm. Assuming $\lambda=60\%$ this gives $F=1.5$ and $\bar{F} \approx 0.53$, an improvement of almost a factor 3. 
This is super-linear in the load (the mean load at look-up being reduced by a factor 2), due to the convexity of $\bar{F}$.

For some comparison of the actual performance of collision rates I have included in Table~\ref{tab:simtable} (Section~\ref{sec:results}) the theoretical and actual collision rate when filling the hash-tables for the four different data sets, and as expected for three of the data sets there are collisions in excess of random, though they are sub-dominant. Unusually, the baryonic simulation has an actual collision rate below that expected for random ($23.7\%$ vs $49.7\%$) - this is likely a symptom of this data set having many contiguous blocks, since the multiplicative hash in Eqns.~(\ref{eq:block_index})~\&~(\ref{eq:hash}) is actually \emph{guaranteed} to give a distinct hash when only a single index is incremented by 1.

One might reasonably wonder if going through the filled blocks in raster order is in fact the optimal approach. Other approaches such as ordering by the index on Peano-Hilbert curve (such as performed in \citealp{Springel_2005}), better preserves spatial locality, which in this context correspond to 
neighbour searches that are more coherent in memory. 
Such schemes introduce the additional complication that the relationship between the values of the 1-d index no longer depends only on the relative positions of the block\footnote{For example a block which is `above' the current (e.g. $i-1$) may have a smaller Hilbert key at one point on the curve, but this will not be true at all points.} and thus a na\"ive implementation requires double the number of neighbour block searches. Examining the various choices of space-filling curve (e.g. Morton/Hilbert) and implementation choices for neighbour searches may produce an interesting direction for future research.

Another avenue for extension is the use of this method on dimensions other than 3. In principle this is straightforward since all of the above methods can be transferred to lower (i.e. 2) or higher dimensions (e.g. the 6-D phase-space FOF of \citealp{Diemand_2006}), although one might need some care since various choices about block-size, hash-function etc. have been calibrated for the 3-D case.

\begin{algorithm}
\caption{Linking of friends of friends groups}
\label{alg:link}
\begin{algorithmic}[1]
\State \emph{\% Precompute indices of blocks reachable from positions 1-64}
\State {\sc Adj}$(p)\gets $ blocks reachable from position $p$
\State \emph{\% Precompute cells reachable within these block pairs $a \in${\sc Adj}$(p)$}
\State {\sc CellReachStencil}$(p_{\rm adj}, p, a) \gets 1$ or 0 (if reachable)
\Procedure{FOF}{$Q, P_{\rm x}, P_{\rm y}, \lambda=0.6$ (desired load)}
\For{all particles ${\bf x}_i$} 
  \State Assign a block $\Phi_i$ \Comment See Eqn.~(\ref{eq:block_index})
  \State Assign position $p_i$ (1-64) of cell within block. 
\EndFor
\State Sort (by blocks and then cell)
\State $B \gets$ hash-table of size $2^n$, with $n$ s.t. $\lambda 2^n>$count(blocks)
\State Create array $C$ to hold cell data
\For{Each block $\Phi_i$ and cell at position $p$ in block $\Phi_i$}
  \State Append cell $c(\Phi_i, p)$ to $C$
  \State Assign parent$(c) \gets c$ (i.e. in own new set)
  \State{\emph{\% Loop over cells already in my block}}
  \For{cell $c_{\rm adj}$ at position $p_{\rm adj}$ already in block $\Phi_i$}
    \State \Call{CompareAndLink}{$c$, $c_{\rm adj}$, $p$, $p_{\rm adj},0$}
  \EndFor

  \State \emph{\% Loop over cells in adjacent blocks}  
  \For{$a$ in {\sc Adj}$(p)$ where $\Phi_i+a \in B$}
    \State $\Phi_{\rm adj} \gets \Phi_i+a$
    \For{cell $c_{\rm adj}$ at position $p_{\rm adj}$ in block $\Phi_{\rm adj}$}
      \State \Call{CompareAndLink}{$c$, $c_{\rm adj}$, $p$, $p_{\rm adj},a$}
    \EndFor
  \EndFor
  \State Append cell $c$ to block $\Phi$
  \State At the last cell, insert $\Phi_i$ into $B$.
\EndFor
\State $\forall c \in C$, assign root($c$) as FOF label for $c$
\State Return .
\EndProcedure
\Procedure{CompareAndLink}{$c_{\rm me}$, $c_{\rm adj}$, $p$, $p_{\rm adj},a$}
  \State \emph{\% If the cells within $b$ and roots are distinct then}
  \State \emph{\% compare points pairwise}
      \If {{\sc CellReachStencil}($p_{\rm adj},p,a$)}

      \If {${\rm root}(c_{\rm adj}) \neq {\rm root}(c_{\rm me})$}
      \If {$\exists \;{\bf x} \in c_{\rm me}, {\bf y} \in c_{\rm adj} : \left| {\bf x} - {\bf y} \right|<b$}
        \State Connect $c_{\rm me}$ and $c_{\rm adj}$ 
        \State using the disjoint sets linking algorithm
      \EndIf
      \EndIf
      \EndIf
\EndProcedure
\end{algorithmic}
\end{algorithm}

\subsection{Periodicity}\label{sec:periodicity}
Periodicity is implemented by the insertion of periodic images around the box in a band of width $b$. For cosmological simulations the number of image points is usually a very small fraction of that of the originals since the linking length is a tiny fraction of the box size.
When such an image point is encountered its cell has its set is assigned to that of the original. 
This does introduce some additional complexity in that when the linking lengths are very large (above a quarter of the box size) the periodic images can be in the same block as the originals and may not be guaranteed to have been inserted yet. This case is covered by falling back on a single-cell block scheme. For large linking lengths it may be preferable to use a scheme where explicit images are avoided and instead the distance function is altered from Euclidean to calculate the distance to the nearest image, however this has not been explored here.

Pseudo-code is described in Algorithm~\ref{alg:link}, although the logic to account for periodicity is omitted for brevity. The interested reader is encouraged to download the source at the aforementioned link. 

\section{Comparison codes and test data}\label{sec:comp}
\subsection{Comparison codes}\label{sec:comp_codes}
The ubiquitous method for performing Friends-of-Friends is the use of a \kdtree.
A \kdtree\ is a method by which the points are recursively partitioned via hyper-planes that are aligned with the coordinate axes at a given point (the median is chosen for a balanced tree), and the axes are either cycled or the longest axis is chosen. Once the tree has been constructed, points within a distance cutoff $b$ can be searched by walking the tree from the root where one either opens or ignores branches depending on the distance criteria, until all neighbouring points are found.

The codes used for comparison are the publicly available \href{https://github.com/N-BodyShop/fof}{FOF} 
of NbodyShop, that is the direct ancestor of current codes such as {\sc cola}\ \citep{Koda_2016}.
During testing the original NbodyShop code occasionally produces inexact group catalogues for large (millions of particles) simulations, and fails entirely for billion particle ones, which turned out to be a result of round-off-error on distances due to the use of 32-bit floating point precision. For comparison a version of this code updated to be double precision (64 bits per value) is used for testing. I believe this is the fairest comparison, since although 32-bit arithmetic is somewhat faster, producing correct group catalogues is in general more important. Timing comparisons have also included that taken to build the \kdtree, since that is also an essential part of the algorithm.

In addition to the NbodyShop code a `dual tree' method \citep[see][for use in correlation functions]{Moore_2001}, which has been applied to the friends-of-friends algorithm by \citet{Feng_2017} is included for comparison. The dual tree method overcomes some limitations of a pure \kdtree\ in very high and low density regimes (where points can be included/excluded branch-wise), though it should be kept in mind that keeping track of the exclusions has some cost in itself, so the comparisons with naive \kdtree\ searches (which are quite light-weight in 3-d) may be marginal, depending on the clustering of the data set. The publicly available version 0.3.27 of {\sc kdcount} has been used, though \citet{Feng_2017} note that the pair enumeration in kdcount was not particularly optimised for performance and so future improvements may be possible.

\subsection{Tests}\label{sec:test_sets}

Table~\ref{tab:simtable} describes the numerical details of the simulations used to test the FOF algorithms. I have performed simulations in both a small volume 10~Mpc box, referred to as `DM fine'  and a larger volume $29.7$~Mpc referred to as `DM coarse', since smaller volumes have slightly more clustering than larger which provides a slightly different challenge. For comparison with larger cosmological volumes, such as simulating large scale structure (LSS, e.g. for weak lensing tests such as  \citealp{Izard_2018}), I have included a $128^3$ DM-only simulation of a 128~Mpc box, referred to as `DM LSS'. 
Since this is a relatively small data set the default timings on this are performed with it tiled twice on each side to make a synthetic $256^3$ data set.

\begin{figure}
\includegraphics[width=0.49\columnwidth]{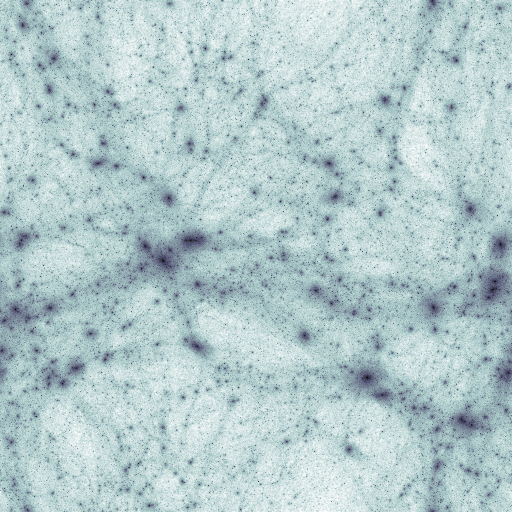}
\includegraphics[width=0.49\columnwidth]{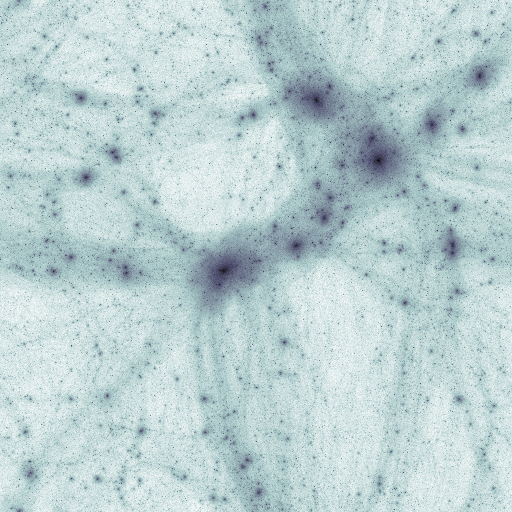}
\includegraphics[width=0.49\columnwidth]{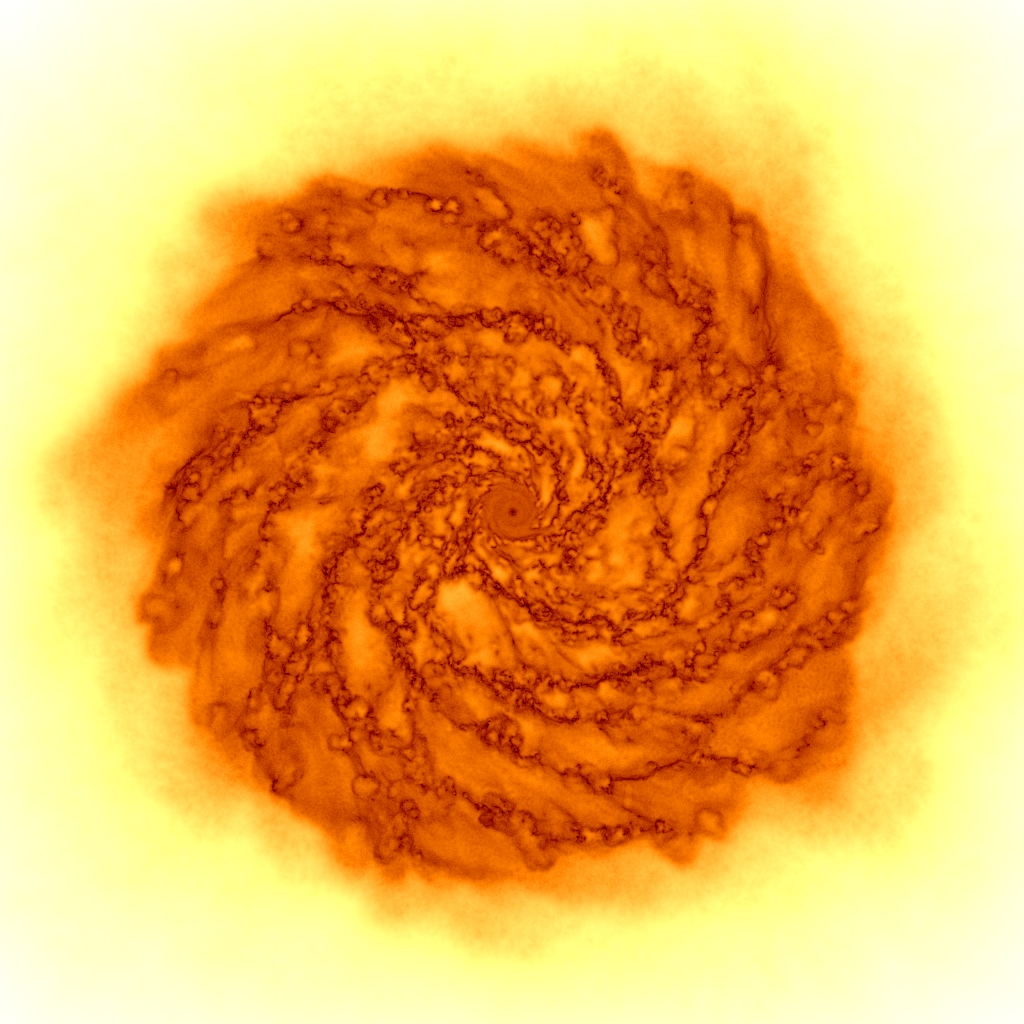}
\includegraphics[width=0.49\columnwidth]{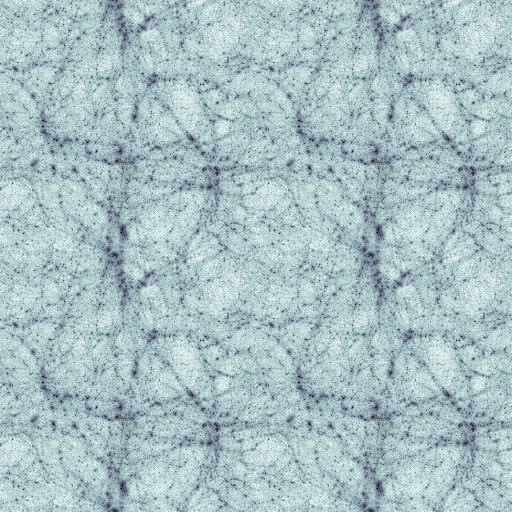}
\caption{Column density plots of simulations used in this work. \emph{Clockwise from top left}, `DM coarse', a $256^3$ dark-matter only simulation of box size $29.7$~Mpc, `DM fine', a $512^3$ dark matter simulation of a $10$~Mpc box used in \citet{Creasey_2018}, 
\emph{below right} `DM LSS', a $128^3$ dark matter simulation of a $128$~Mpc box, tiled 8 times to make a $256$~Mpc box and 
\emph{below left} `Baryonic', a gas distribution of a MW-like galaxy with 1,651,651 gas particles (in Hicks and Sales et al., in prep.)}
\label{fig:sims}
\end{figure}

To broaden the tests a baryonic simulation of a galactic disk is included. Here the FOF algorithm can be used for example for the identification of clumps in the interstellar medium (Hicks and Sales, in prep.), and as such the gas particles from up to $20$~kpc from the galactic centre have been considered. A useful density threshold for this kind of simulation is that of star formation at around $0.1\;m_{\rm p}\, {\rm cm}^{-3}$ (where $m_{\rm p}$ refers to the proton mass), which corresponds to a linking length of 237~pc at this resolution. Projected images of these simulations are shown for an overview in Fig.~\ref{fig:sims}.

A number of operations in this work are sensitive to the \emph{initial} order in which the particles data is stored, in particular the index to sort the data in raster order is quicker to construct for locally coherent positions, and the direct distance comparisons also benefit from cache `hits' when the data is coherent.
Some codes, for example {\sc Arepo}, will by default order their particle outputs according to their friends of friends groups. The process of group-finding in an already group-sorted list is by some measure `cheating' (provided you want the exact same linking length), and so in the fiducial tests the points are sorted into Morton ordering before attempting a group-find. For completeness some tests are performed with a random ordering, though this is more of a `worst case', since every simulation code I know of orders its outputs in some spatially coherent way.

Finally, to give us some additional simulation sizes to compare for scaling, all the DM simulations are tiled in $x$,$y$,$z$ (increasing the particle counts in powers of 8) to make artificial simulation data sets up to $1024^3$ points.

All the benchmarks shown in the figures and table in section~\ref{sec:results} have been performed on a node of a machine using a single core of an AMD 6380 Opteron\texttrademark\ $2.5$~GHz processor with a 16~MB L3 cache and 256~GB of RAM, though in order to mitigate `over-fitting' to a particular CPU architecture I have also tested some of the smaller data sets on an Apple iMac with a 4-core $3.2$~GHz Intel Core i5 processor (6~MB cache, 8GB RAM) to verify that the ranking of the optimisations is unchanged. Each test was performed three times and a `best of three' timing is given, except for the $1024^3$ data sets on the tree-based codes, where only one test was run due to time constraints. 

\begin{table*}
\begin{center} 

\begin{tabular}{lrrrr}
Simulation & DM coarse & DM fine & Baryonic & DM LSS\\
\hline
Region width & $29.7$ Mpc & 10 Mpc & 40 kpc & $2\times128$ Mpc\\ 
Points in volume  & $256^3$ & $512^3$ & 1,651,651 & $8\times128^3$ \\
Periodic images & 0.24\% & 0.048\% & Isolated & 0.18\% \\
Av. points/filled cell & 1.5 & 1.6  & 6.0 & $1.25$ \\
Av. points/filled block &  3.0 & 3.9 & 86.8 & $2.1$ \\
Linking length & $b=0.2$ & $b=0.2$ & 237 pc & $b=0.2$\\
$\;$ corresponds to & 200$\rho_{\rm crit}$& 200$\rho_{\rm crit}$ &  $0.1 \; m_{\rm p} {\rm cm^{-3}}$ & 200$\rho_{\rm crit}$\\
\# linked groups  & 5,983,049 & 37,140,510 & 17,283 & 7,975,232 \\
Fill collision rate &  22.3\% & 50.4\% &  23.7\% & 40.2\% \\
c.f. random rate $F(\lambda)$ & 21.7\% &  40.9\% &  49.7\% &  34.9\%\\
\hline
 Timings (s) & & & \\
hfof  & 10.26 & 96.68 & 0.39 & 7.62 \\
$\;$ (sort) & (2.77) & (28.84) & (0.14) & (2.04) \\
\hline
NbodyShop & 75.66 & 1112.63 & 19.23 & 39.18 \\
$\;$ (make \kdtree) & (10.39) & (138.77) & (0.72) & (10.39)\\
\hline
kdcount & 124.45 & 1693.17 & 16.42 & 72.81 \\
\hline
\end{tabular}

\end{center}
\caption{Simulation details for four data sets used in this paper. Boldface values indicate the total time for a code, values in parentheses are partial. Illustrations of these simulations can be seen in Fig.~\ref{fig:sims}.}\label{tab:simtable} 
\end{table*}

\section{Results}
\label{sec:results}

\begin{figure*}
\includegraphics[width=\columnwidth]{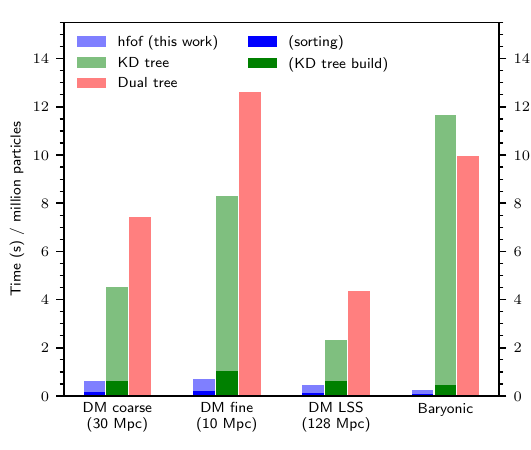}
\includegraphics[width=\columnwidth]{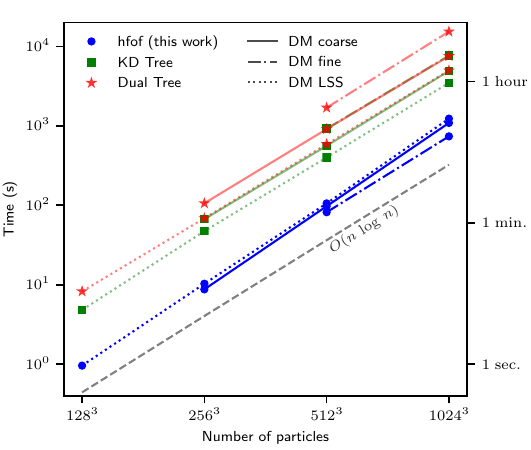}
\caption{ Timing for FOF calculations at different scales. \emph{Left panel}: Time per million particles for the simulations in Table~\ref{tab:simtable}. \emph{Blue bars} shows \hfof\ (this work), with the \emph{dark blue region} showing the time spent just on the sort (around a $30\%$ of the total). \emph{Green bars} indicates FOF using neighbour searches in a \kdtree\ (NbodyShop), with \emph{dark green region} indicating just the \kdtree\ construction. \emph{Red bars} shows the dual-tree method (of {\sc kdcount}). 
\emph{Right panel}: Total time (seconds) when the periodic DM simulations are tiled to make larger data sets. 
\emph{Solid lines} indicate the DM~coarse simulation, \emph{dot-dashed} the DM~fine simulation and \emph{dotted line} the DM~LSS simulation, all tiled in factors of 8 up to $1024^3$.  
\hfof\ is in \emph{blue circles}, the KD tree in \emph{green squares} and the dual tree in \emph{red stars}, with colours as per left panel. 
\emph{Grey-dashed line} indicates the comparison growth scaling of $O(n \log n)$. Each DM-only simulation used the fiducial linking length of $0.2\times$ the interparticle spacing.}
\label{fig:time_fracs_scaling}
\end{figure*}

In Fig.~\ref{fig:time_fracs_scaling} the time taken to perform a FOF group finding exercise is plotted per million particles for each of the simulations shown in Table~\ref{tab:simtable}. In the right panel the DM simulations are `tiled' (repeated in $x$,$y$,$z$) to make higher particle counts all the way up to $1024^3$ to examine the scaling.
As expected, there is an almost $O(n \log n)$ scaling with number of particles, comparable to the \kdtree\ based codes. 
The timings of \hfof\ are around an order of magnitude faster, with the \kdtree\ codes performing better for the largest volume DM simulation (DM~LSS) and worse for the smallest volume DM~fine $512^3$ simulation, where \hfof\ is nearer a factor 20 faster than kdcount, indicating that \hfof\ scales more weakly with clustering.
The `Baryonic' simulation has a relatively small number of groups (i.e. highly clustered, see Table~\ref{tab:simtable}), and the dual tree method does better than the \kdtree.
Notably {\sc hfof} (this work) spends around a third of its time sorting, whilst the \kdtree\ codes spend more time constructing the \kdtree\ itself than is spent on the entire FOF search, at all simulation sizes. It is emphasised that all the algorithms here produce \emph{exactly} the same friends-of-friends groups. This is verified by re-labelling the FOF groups by the lowest index of the points they contain (in the original point list), and checking that each point is assigned to a FOF group with the same label.

\begin{figure}
\includegraphics[width=\columnwidth]{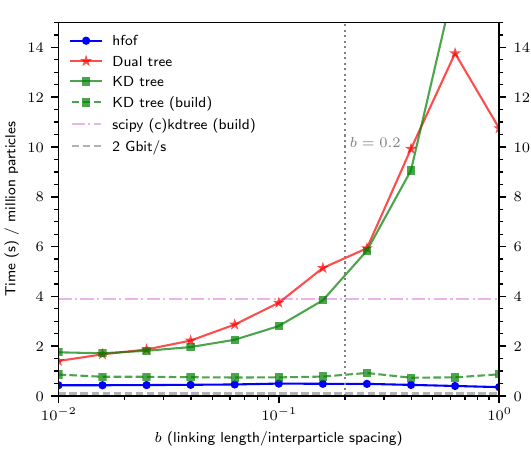}
\caption{Time to calculate groups as a function of linking length for the DM coarse ($256^3$) simulation. Time per million particles as a function of $b$, the linking length per interparticle spacing, where timings are taken from a `best-of-three' for each code and linking length. \emph{Solid lines and symbols} are coloured as for the right panel of Fig.~\ref{fig:time_fracs_scaling}. The fiducial $b=0.2$ of \citet{Davis_1985} is indicated by the \emph{vertical grey dotted line}. For comparison the \emph{faint magenta dot-dashed line} indicates the construction time of scipy's \kdtree\ whilst the \emph{horizontal grey dashed line} indicates an estimate of the sequential read speed of the file system used (around 10 million particles per second). }
\label{fig:time_spacing}
\end{figure}

To inspect the dependence on linking length, Fig.~\ref{fig:time_spacing} shows the timing variations for the DM coarse simulation of linking lengths from $0.01$-$1\times$ the interparticle spacing. In order to reduce the stochasticity due to background processes, all timings for this have been repeated and a `best-of-three' timing has been given (periodicity has also been disabled for this calculation). We see that at smaller linking lengths the \kdtree\ based codes spend around the same amount of time at the linking stage as the \kdtree\ construction, a fraction which steadily drops to larger linking lengths. At a ratio of $0.2\times$ the interparticle spacing the fraction is around 20\%.
For comparison the \kdtree\ construction time for the popular \href{https://docs.scipy.org/doc/scipy/reference/generated/scipy.spatial.cKDTree.html}{scipy.cKDTree} is\\
included for the same data sets, which is significantly slower than that of the NbodyShop - likely because the latter has been specialised to the 3d case. 
The trend of the spatial hashing method to have a comparable or faster completion time to the \kdtree\ construction time (which is independent of linking length) appears to be true at all $b$.
At extremely short linking lengths, where the problem is essentially one of checking points are isolated, it is around $3\times$ faster, rising to above $30\times$ for the largest (most connected) linking lengths. This scaling with connectivity is the likely reason for the trend with volume/clustering in Fig.~\ref{fig:time_fracs_scaling}, since the 10~Mpc simulation has more connectivity (i.e more particles per group) than the larger box. 
For the largest linking length ($b=1$), {\sc kdcount} is more efficient than it is at $b=0.91$ (see last 2 points on the red line) which is consistent upon repeated testing. This improvement is likely due to their focus on fast merge operations, and is at least visually consistent with \citet[red line in the left panel of their Fig. 6]{Feng_2017}. 

These rates are high enough that loading the simulations from the disk starts to become non-negligible concern, and Fig.~\ref{fig:time_spacing} includes an estimate of the read-speed from the file-system used, around 2~Gbit/s or 10~million points per second. An unintended consequence of this speed is that the algorithm may be difficult to efficiently parallelise on distributed memory architectures, since the single process calculation can be completed at speeds that are likely within a factor of a few of the Ethernet speed, though certainly an implementation could be constructed where the total volume was decomposed into per-process sub-volumes, each with their own hash-table and images of the boundary points.

\begin{figure}
\includegraphics[width=\columnwidth]{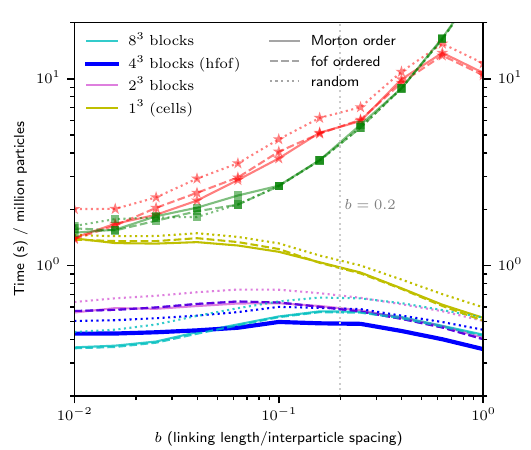}
\caption{Speed as a function of block size for the DM coarse simulation (no periodic images) with $8^3$~cells in \emph{cyan}, $4^3$ (\hfof) in \emph{blue}, $2^3$~in magenta and $1^3$ (i.e. cells only) in yellow. Different line styles indicate effects of the point orderings, with Morton ordering in \emph{solid}, FOF ordered in \emph{dashed} and random in \emph{dotted}. \emph{Green squares} and \emph{red stars} show timings with the \kdtree\ based codes as in Fig.~\ref{fig:time_spacing}.}
\label{fig:speed_blocks}
\end{figure}

In Fig.~\ref{fig:speed_blocks} the effects of changing the block size and of different point orderings on the time for FOF on the DM coarse simulation ($256^3$) are shown. Periodicity has not been implemented for all block sizes, and so all algorithms are tested here without periodic images. The point orderings shown are those ordered by FOF group (recall that \hfof\ will then re-index into raster order), ordering by Morton index, and ordering randomly. As one might expect, having the points ordered by their FOF groups is generally the fastest (since they are essentially pre-grouped), although using a Morton ordering generally shows timings within 5\%. Having positions randomly ordered is definitely the most difficult case, though as mentioned in Section~\ref{sec:test_sets} this is not a normal situation.

With respect to the block sizes it appears that using larger block sizes improves performance for very short linking lengths. Presumably this is because when you only have to check isolation, having a greater fraction of the neighbour cells in the same block (i.e. adjacent memory location) is still preferable. 
For example in section~\ref{sec:cells} we saw that with $1^3$ blocks (i.e. individual cells) an average of 58 neighbouring blocks would be loaded, rather than $3.4$ with $4^3$ blocks. As the linking lengths are increased, smaller blocks become more competitive, larger blocks having to iterate over spurious cells that cannot be neighbours (they are outside the mask). 
Using $4^3$ blocks seems to be a good compromise, in that it appears the fastest at higher linking lengths (probably partially since the current processors have architectures well suited to the storage of 64-bit masks for use in cell exclusions based on positions), and even at the shortest linking lengths is only around $20\%$ slower than using the larger $8^3$ blocks.

\section{Discussion}\label{sec:conc}
This work describes the implementation of a spatial hashing algorithm for friends-of-friends group identification in 3-d. 
The algorithm groups points into cells whose extent is bounded by the linking length, and blocks of these cells are spatially hashed for fast neighbour searches, the cells being incrementally merged via the disjoint sets algorithm.

The conceptual novelty of this approach when compared to other FOF algorithms is that the ($k$-d) tree structure for the particle data has been entirely dispensed with, replaced instead with spatial hashing for fixed-distance neighbour look-ups. With the addition of other optimisations such as the grouping of cells into blocks (to avoid excessive memory look-ups) and exploiting the symmetry of the distance metric to avoid double counting and partially fill the hash-table leads to an algorithm that typically completes in less time than a \kdtree\ construction. 

This code has been tested on cosmological simulations of up to a billion particles, and a baryonic (galaxy) simulation of over a million particles. The numerical results demonstrate speed increases of up to $50\times$ in the baryonic case and up to $20\times$ in the cosmological, with $8\times$ a more representative improvement. 

This work has focused on the 3-dimensional case, however all of the techniques here are applicable in $k$-dimensions, and future work could include the application of these particularly for 2-dimensions. It may also be possible to eliminate the particle sort, although that was not the dominant computational expense, and parallelisation of the algorithm could be explored.

\section*{Acknowledgements}
PEC would like to thank Lars Hernquist, Ben Lowing, Laura Sales and Anson D'Aloisio for interesting discussions about spatial hashing and useful comments. PEC would also like to Y. Feng and C. Modi for making the {\sc kdcount}\ method open-source and painless to install, and the anonymous referees for helpful comments which improved this manuscript.
All the simulations in this paper were performed on Raptor, a cluster using AMD 6300 series Opteron\texttrademark\ processors.

\section*{References}
\bibliography{paper}

\end{document}